# New evidence for determining of the date of adoption of Christianity as a state religion in Georgia

MARINA GIGOLASHVILI, ROLAN KILADZE, GEORGE RAMISHVILI
VASILI KUKHIANIDZE
Georgian E. Kharadze National Astrophysical Observatory at
Ilia Chavchavadze State University, Tbilisi, Georgia

### 1. Introduction

At present it is considered that Christianity was declared a state religion in Georgia in circa 326 AD, during the reign of King Mirian and Queen Nana. In the Georgian Chronicle [1] (Kaukhchishvili, 1955) it is stated that this event is connected with the adoption of Christianity by King Mirian. Once he was hunting somewhere between Mtskheta (the ancient capital of Georgia) and Khashuri, near Mt. Tkhoti in dense woodland. It rapidly got dark and the Sun disappeared from the sky.

Mirian began to ask his traditional pagan gods, but to no avail. Then he addressed the god whom Nino from Cappadocia believed in (subsequently she became Saint Nino, a woman whose name is inseparably linked with the spread of Christianity in Georgia) and there was a miracle; the darkness suddenly disappeared and the Sun began shining in the sky again. Then Mirian turned to the East and thanked "Nino's god".

### 2. Contemporary investigations

In the 1930s the Georgian historian Ivane Javakhishvili appealed to astronomers to answer the question whether a total solar eclipse happened in Georgia in the fourth century or not. Based on the famous *Canon der Finsternisse* of Oppolzer [2], in which the change in the length of the day with current time (the result of tidal friction) was taken into account incompletely, astronomers could not find any eclipse during the mentioned period. Thus it seemed that the question was resolved.

However, after detailed tables and maps of solar and lunar eclipses had been published on the Web by Espenak [3], we have found out that a total solar eclipse did indeed occur in Georgia on 6 May 319 AD.

### 3. The circumstances of the solar eclipse

By calculating the circumstances of the eclipse with the use of Bessel's improved elements, we have found that Mount Tkhoti was on the central line of the eclipse [4-6]. For the place where Mirian was hunting ($\lambda = 44°.55$; $\varphi = +41°.99$), the circumstances of the eclipse are as follows: the start of the partial eclipse was at $14^h58^m01^s$ Universal Time (UT); the second contact was at $15^h51^m57^s$ UT; the third contact was at $15^h53^m50^s$ UT; the maximal phase was 1.018. The moments of sunset are as follows: bottom edge - at $15^h59^m24^s$ UT, top edge - at $16^h02^m29^s$ UT.

The central line of the eclipse passed through the settlements of Tsageri, Ambrolauri, Tskhinvali and Mtskheta. The northern boundary passed through the Caucasus Range (Elbrus, Upper Baksan and Kazbek). The southern boundary passed through Lake Paliastomi, Abastumani, Aspindza, Dmanisi and Akhtala. From the east the strip of the complete eclipse was limited by a line from Gardabani to Sagarejo.

Hence, the eclipse happened in the evening, before sunset; the duration of the total phase was about 2 min. At the moment of the maximal phase the height of the Sun above horizon was only $0°.8$. The sunset began 5.6 min later, after the end of total phase (i.e., after the third contact).

We have investigated every solar eclipse (total, partial and annular) during the period 290-365 AD. In Table 1, the list of solar eclipses with a phase more than 0.8 for the period mentioned above is given for Mt. Tkhoti. In the columns of Table 1 are consistently given: the year, the month and the day of an eclipse, the moments of the first and second contacts, the maximal phase and the third and fourth contacts. In the last two columns are given the heights of the Sun above horizon (in degrees) at the moment of the maximal phase of an eclipse and the maximal phase (in %).

Table 1. The list of solar eclipses with a phase more than 0.8 during the reign of King Mirian

| Year | Mn | Day | 1 cont. | 2 cont. | Max. ph. | 3 cont. | 4 cont. | Altitude | Phase |
|---|---|---|---|---|---|---|---|---|---|
| 306 | 7 | 27 | $7^h17^m34^s$ | | $8^h39^m49^s$ | | $10^h15^m05^s$ | 41°.4 | 83.2 |
| 319 | 5 | 6 | 17 57 58 | $18^h51^m54^s$ | 18 52 50 | $18^h53^m47^s$ | | 0.9 | 101.7 |
| 346 | 6 | 6 | 6 18 52 | | 7 16 42 | | 8 19 44 | 29.6 | 99.3 |
| 348 | 10 | 9 | 7 40 54 | | 8 45 24 | | 9 54 31 | 25.7 | 87.6 |
| 355 | 5 | 28 | 5 55 51 | | 6 53 35 | | 7 56 47 | 24.7 | 87.3 |

**4. Other evidences**

According to *Kartlis Tskhovreba* [1]), 3 crosses from cypress were made on 1 May. According to Ioane Zosime [7], it was the third Sunday after Easter. One of these crosses was raised near the capital of Georgia, Mtskheta, on 8 May. However, the year is not mentioned in these sources.

From further investigation it will be clarified that the event studied by us occurred before the First Council of Nicaea (325 AD). Hence, the contemporary rules of calculation on Easter were not still canonized. For this reason the authors have calculated the date on Easter by all possible methods.

To ascertain the exact date of this incident the authors investigated the data on all Easters during the probable period of the reign of King Mirian. With this period, the years when Easter took place on 17 April, have been chosen. Hence, the third Sunday after Easter fell on 1 May.

In Table 2, the data on Easters in the appropriate years are presented. In the first 3 columns are given the moments (year, month, day, hour and minute) of the first full moon after a spring equinox, according to Espenak [2]; in the following columns the data on Easter calculated by us by using different methods are given: by the ancient 19-year lunar cycle and approximate formulas of Gauss and Meeus [8-10]. In the last column of the table are given Easter data, selected by the authors with the use of the exact moments of full moon.

Table 2. The data on Easters during the reign of King Mirian

| | Espenak | | Lunar Cycle | Gauss and Meeus | Selected data |
|---|---|---|---|---|---|
| 298 | 13 Apr | $23^h56^m$ | 13 Apr | 17 Apr | |
| 309 | 11 Apr | 14 15 | 12 Apr | 17 Apr | |
| 315 | 6 Apr | 11 28 | 5 Apr | 10 Apr | |
| 320 | 9 Apr | 14 01 | 10 Apr | 10 Apr | 17 Apr |
| 326 | 4 Apr | 4 47 | 4 Apr | 3 Apr | 10 Apr |
| 337 | 1 Apr | 23 23 | 2 Apr | 3 Apr | |
| 343 | 27 Mar | 19 23 | 27 Mar | 27 Mar | 3 Apr |
| 348 | 31 Mar | 13 13 | 1 Apr | 3 Apr | |
| 354 | 25 Mar | 10 22 | 25 Mar | 27 Mar | |
| 365 | 23 Mar | 20 14 | 24 Mar | 27 Mar | |

As is clear from Table 2, a 17 April Easter could only take place in the years 298, 309 and 320 AD. In Table 1, only two eclipses (306 and 319) are presented which could happen before the years when Easter took place on 17 April. However, the eclipse of 306 must be excluded for two reasons: it was a partial eclipse (with a maximal phase of 82%) and happened early in the morning. But this incident happened to King Mirian in the evening [1].

## 6. Conclusions

In the authors' opinion, the eclipse seen by King Mirian happened on the evening of 6 May, 319 AD. Later, in May 320 AD, cypress crosses were made and raised. Thus, we have answered the question raised 70 years ago by the Georgian historian Ivane Javakhishvili about the occurrence of a total solar eclipse in Georgia in the fourth century. The eclipse seen by King Mirian happened on the evening of 6 May, 319 AD. Later, in May 320 AD, cypress crosses were raised and Christianity was become the state religion of Georgia.